# All fiber ultrafast laser generating gigahertz pulse based on a hybrid plasmonic microfiber resonator

Zi-xuan Ding[1], Zi-nan Huang[2], Ye Chen[1], Cheng-bo Mou[2], Yan-qing Lu[1,*] and Fei Xu[1,*]

[1]College of Engineering and Applied Sciences, Nanjing University, Nanjing 210093, China

[2]Key Laboratory of Specialty Fiber Optics and Optical Access Networks Shanghai, Shanghai University, Shanghai 200072, China

*feixu@nju.edu.cn and yqlu@nju.edu.cn

Abstract

Ultrafast lasers generating high repetition rate ultrashort pulses through various mode-locking methods can benefit many important applications including communication, materials processing, astronomical observation, etc. For decades, mode-locking based on dissipative four-wave-mixing (DFWM) has been fundamental in producing pulses with repetition rates on the order of gigahertz (GHz), where multiwavelength comb filters and long nonlinear components are elemental. Recently, this method has been improved using filter-driven DFWM, which exploits both the filtering and nonlinear features of silica microring resonators. However, the fabrication complexity and coupling loss between waveguides and fibers are problematics. In this study, we demonstrate a tens to hundreds of gigahertz stable pulsed all-fiber laser based on the hybrid plasmonic microfiber knot resonator device. Unlike previously reported pulse generation mechanisms, the operation utilizes the nonlinear-polarization-rotation (NPR) effect introduced by the polarization-dependent feature of the device to increase intracavity power for boosting DFWM mode-locking, which we term NPR-stimulated DFWM. The easily-fabricated versatile device acts as a polarizer, comb filter, and nonlinear component simultaneously, thereby introducing a novel application of microfiber resonator devices in ultrafast and nonlinear photonics. We believe that our work underpins a significant improvement in achieving practical low-cost ultrafast light sources.

**Introduction**

High-repetition-rate pulsed lasers have garnered considerable attention during recent decades for applications in optical communications [1], microwave photonics [2], generation of frequency combs [3] and spectroscopy [4]. However, as the most extensively used pulsed laser source, mode-locked fiber laser typically produces pulse trains with fundamental repetition rates well below 1 GHz owing to the limitations of the laser cavity's length. Therefore, several methods for enhancing repetition rates have been proposed, including direct approaches such as active mode-locking [5] or shortening of cavity length [6], although the repetition rates were still below 20 GHz [7] with degraded pulse quality due to a decline in intracavity energy. Passive harmonic mode-locking methods proposed subsequently increase the repetition rates of pulsed lasers up to tens of gigahertz by producing multiple pulses in each round trip [8, 9], while the harmonic mode-locking operation is hardly controllable.

In 1997, Yoshida et al. demonstrated a new approach for generating 115 GHz pulse trains, in which a Fabry-Pérot filter and 1.5 km dispersion-shifted fiber were incorporated in the main cavity [10]. The underline mechanism is called dissipative four-wave-mixing (DFWM) [11], whose key components comprise both multiwavelength filters and high-nonlinear elements. Since then, a number of demonstrations of high repetition-rate pulse trains adopting such a method have been reported, exploiting various devices such as fiber Bragg gratings [12], MZ interferometers [13], and silicon microring resonators [14]. In 2012, Peccianti et al. proposed a stable 200 GHz ultrafast fiber laser based on a silica microring resonator. The resonator serves as both a comb filter and high-nonlinear element to boost DFWM mode locking [15], thus they termed the mechanization filter-driven FWM (FD-DFWM). Nevertheless, the silicon/silica scheme often involves high cost and significant coupling loss between the fiber and the silicon/silica-waveguide. Therefore, a low-cost all-fiber resonator for generating high repetition-rate pulse fiber lasers using DFWM is desired. Nevertheless, strong nonlinearity is required to trigger short pulse generation.

Owing to advantages of strong evanescent field, low insertion loss as well as compatibility to all-fiber optical systems, microfiber-based devices have been widely used, especially for microfiber resonators [16, 17, 18].It is noteworthy that with significantly smaller diameters and air cladding, tapered microfibers exhibit high nonlinearity compared with common single-mode fibers (SMFs); a 2 μm microfiber's nonlinear coefficient γ is calculated to be approximately 50 $W^{-1}$/km at 1550 nm [19], which is 50 manifold that of the standard SMF, although still lower than silicon\silica waveguides'110-220 $W^{-1}$/km [15, 20]. Hence microfiber can be combined with two-dimensional materials for nonlinearity enhancement [21]; however, saturable absorption and additional insertion losses would be introduced.

Herein, we report a ring fiber laser incorporating a hybrid plasmonic microfiber knot resonator (HPMKR) applied to generate pulses up to 144.3 GHz at 1550 nm. The large polarization dependent loss (PDL) of the HPMKR results in the nonlinear polarization rotation (NPR) of the laser cavity, yielding a Q-switched or mode-locked pulse with large instantaneous power to compensate relatively low nonlinearities and excite DFWM in the microfiber. For its versatile role in fiber lasers we tend to term the laser scheme as NPR-stimulated DFWM. The salient point is that the HPMKR is not only a broadband polarizing element, but also a filter. The laser oscillates in stark contrast to all previous DFWM schemes, where the necessity of extremely high nonlinear elements is removed. In addition, the complexity hindering high Q (million) device fabrication may be removed, thereby lowering the bar for achieving DFWM effectively.

## Results
### Fabrication and polarization features of HPMKR

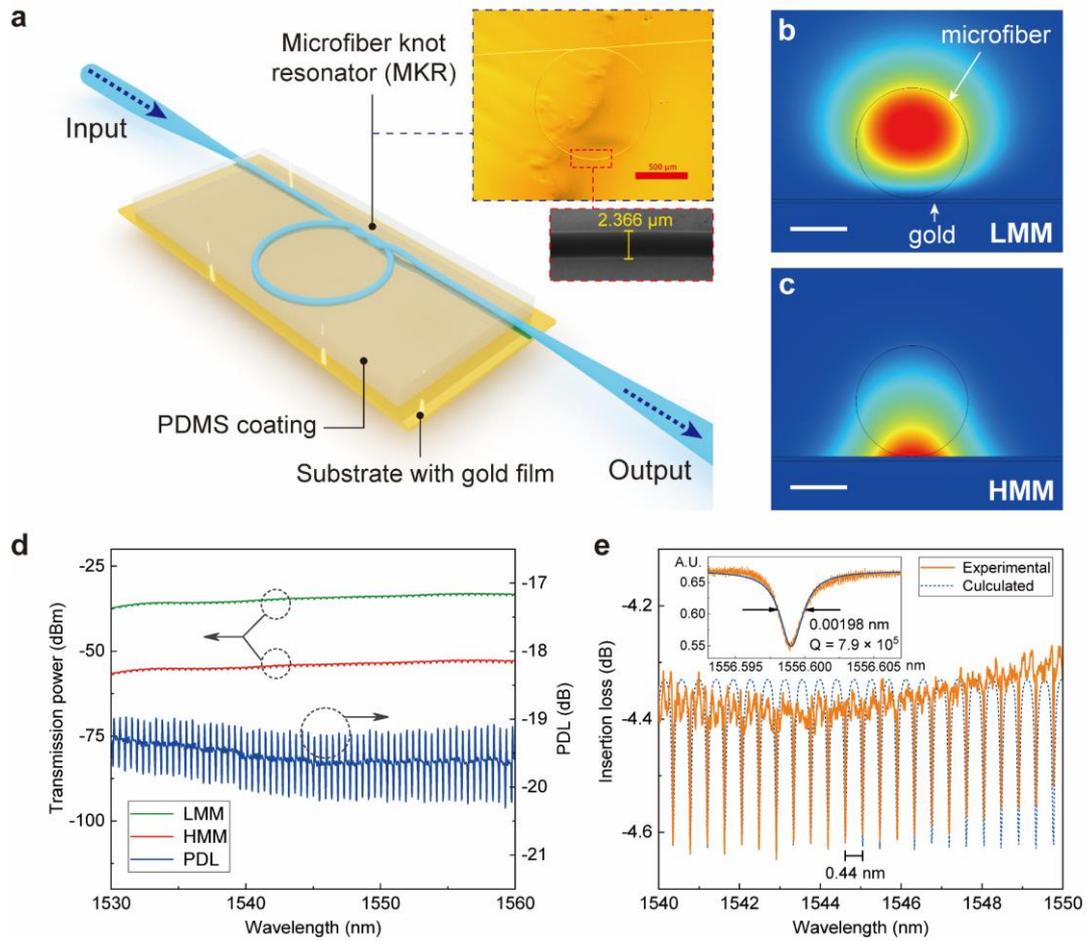

Figure 1 | HPMKR design and characterization.

The key device of the laser, HPMKR, shown in **Fig. 1a** exhibits a concise structure. A knot resonator formed from tapered microfiber was attached to a glass substrate with a gilded surface and then packaged with polydimethylsiloxane (PDMS) polymer (Specific fabrication process provided in Methods). After the complete solidification of PDMS, the HPMKR turns into a stable device with steady properties to be characterized. As for the ring resonator, the resonant condition can be expressed as $n_{eff} \times L = m\lambda_{res}$, where $n_{eff}$ is the effective refractive index, $L$ the ring section's length of MKR, $\lambda_{res}$ the resonant wavelength, and $m$ an integer. From this equation and the coupled mode theory, the intensity transmittance of the MKR can

be derived as [22]

$$T = (1-\gamma)\left\{1 - \frac{[1-(1-\gamma)\exp(-\rho L)\sin^2(\kappa l)]}{\left[1-(1-\gamma)^{1/2}\exp\left(-\frac{\rho}{2}L\right)\cos(\kappa l)\right]^2 + 4(1-\gamma)^{1/2}\exp\left(-\frac{\rho}{2}L\right)\cos(\kappa l)\sin^2\left(\frac{\beta L}{2}\right)}\right\}$$

where $\kappa$, $l$, $\gamma$, and $\rho$ denote the mode-coupling coefficient of the coupling region, coupling length, intensity-insertion loss coefficient and intensity attenuation coefficient of the MKR, respectively. To experimentally verify the spectral response of the HPMKR, ASE light at 1550 nm was injected into the device; **Fig. 1e** shows a typical transmission spectrum of the hybrid device, where the Q-factor is measured to be 7.9 × $10^5$ (see methods for Q-factor measuring details), which is high for an MKR.

Considering the strong SPP introduced by gold, certain polarization-dependent features should appear owing to the structure's lack of circular symmetry. As the ring section establishes contact with the metal surface, the coupling between the plasmonic mode supported by the gold film and the waveguide modes results in a hybrid plasmonic mode (transverse-magnetic-like mode) [23]. The high-metal-coupling-loss mode (HMM) and low-metal-coupling-loss mode (LMM) will exhibit different losses during propagation in the HPMKR. Using the finite element method, we can numerically investigate the polarization-related mechanism; **Fig. 1b,c** illustrates the cross-section electric field distributions of both modes. The diameter of the microfiber is 2.5 μm, the thickness of gold film 100 nm, the refractive indexes of gold and PDMS 0.5582+10.756i and 1.3997, respectively (both at the wavelength of 1550 nm).

Next, we measured the polarization dependent loss (PDL) of the device. Light from an ASE

light source operating at 1550 nm was linearly polarized by a polarizer; subsequently a polarization controller (PC) was used to adjust the polarization state of light injected into the HPMKR, in which different optical modes in the HPMKR can be excited selectively. An optical spectrum analyzer (OSA, Yokogawa, AQ6370C) was employed to record the transmission spectrum. **Fig. 1d** shows the largest PDL of HPMKR sample that we measured experimentally, whose value reaches up to 19.75 dB at 1550 nm, implying the device's great potential as a polarizer. For comparison, a commercial polarization-dependent isolator usually has a PDL of ⩾ 20 dB (IO-G-1550-APC, Thorlabs, U.S.).

**Laser scheme and operation**

The experimental set-up is shown in **Fig. 2**. The HPMKR sample was embedded in a standard erbium-doped fiber (EDF) ring laser cavity. The EDF (OFS RightWave® EDF80) features a dispersion of -48 ps/(nm·km), providing active gain for operation at 1550-nm waveband. The fibers comprising the remaining cavity are all SMFs. The length of the EDF and SMF can be adjusted to investigate the laser's operation status under different net dispersions. A polarization-independent isolator (PI-ISO) was employed to force the pulse to circulate unidirectionally, and two polarization controllers act on the pulse polarization state as the HPMKR is a polarization-sensitive device. The waveform at the 20% output of the optical coupler (OC) is monitored with an autocorrelator (A.P.E, pulseCheck-50) to measure the pulse duration, in addition to an OSA and a 200 MHz photodetector connected with an oscilloscope and radio frequency spectrum analyzer.

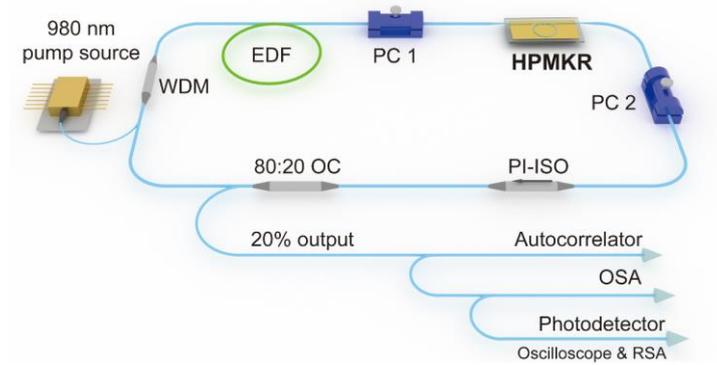

Figure 2 | Schematic of the fiber laser based on HPMKR.

Figure 3 shows the optical spectra (**Figs. 3a,a-1,a-2**) and the time-domain traces of the pulsed output at different positions of PCs obtained by an oscilloscope (**Figs. 3b,b-1,b-2**) and an autocorrelator (**Figs. 3c,c-1,c-2**). Here the laser cavity composed of a 1.7 m EDF and 7.87 m SMF possessed a net dispersion of -0.075 $ps^2$, while the total length of 9.57 m results in a basic longitudinal mode spacing of 21.6 MHz. The sample used here is labeled sample-A; It exhibits an FSR of 0.44 nm (corresponding to 54.1 GHz at 1550 nm) and a high-Q-factor of 786,161, as shown by the transmission spectra in **Fig. 1e**. The insertion loss of this sample is 4.27 dB and the PDL is measured to be 6.93 dB at 1550 nm, which has been experimentally proven to be sufficient to drive NPR.

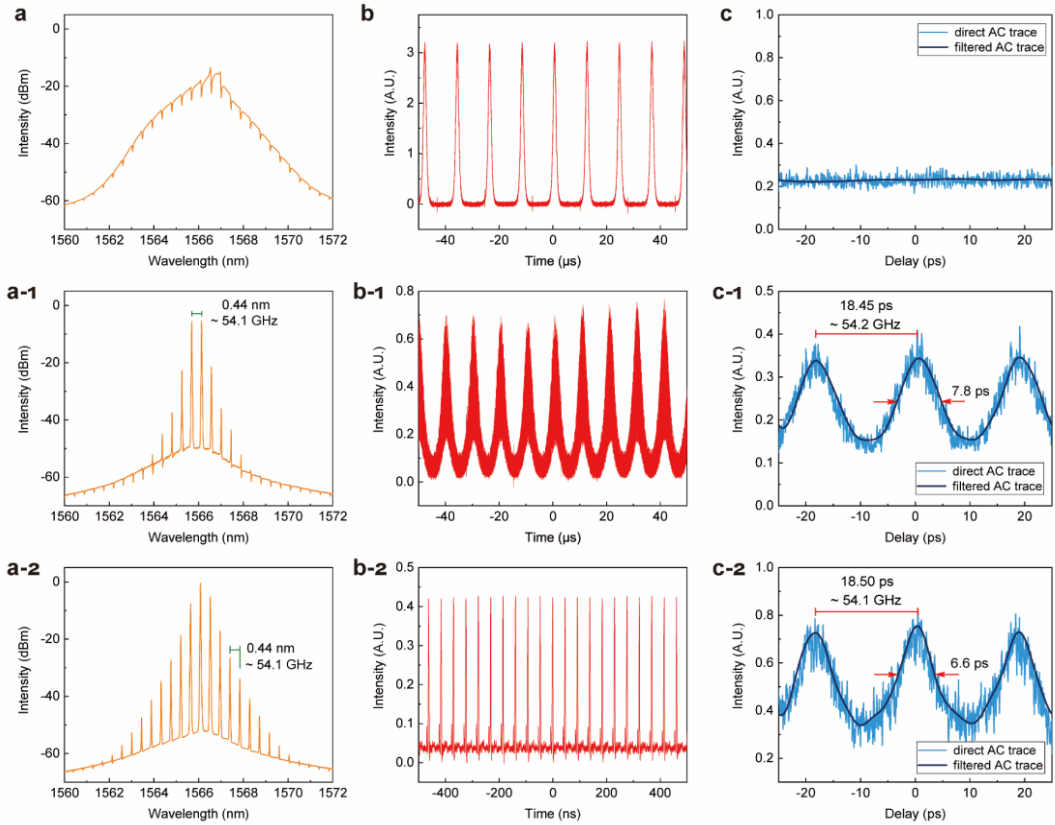

**Figure 3 | Evolution process of NPR-stimulated DFWM operation.**

The laser started lasing when the pump power reached approximately 35 mW; however, only a single-wavelength operation was observed under this circumstance. When the pump power was increased to ~100 mW, with fine tuning of PCs, the Q-switched pulse train was monitored by the oscilloscope, which is a typical consequence of the NPR effect. To further stabilize the pulse, the pump power was increased to 390 mW, where a stable Q-switched pulse train was achieved (**Fig. 3b**). The corresponding optical spectra shown in **Fig. 3a** exhibit a Q-switched signal's spectra profile that was filtered by MKR. The autocorrelator detected only DC signals as shown in **Fig. 3c**. By slightly rotating the PC, the output optical spectra evolved into a multiple-wavelength profile plot in **Fig. 3a-1**. All lasing wavelengths corresponds well to the HPMKR's resonances centered around 1566 nm with a uniform interval of 0.44 nm, which was defined by the sample's FSR. Meanwhile, the temporal trace recorded by oscilloscope became a Q-switched mode-locking pulse train with peak intensity relatively declining by 4.6 times (**Fig. 3b-1**). However simultaneously the autocorrelator gave a high-repetition-rate pulse train with a pulse-to-pulse interval of 18.45 ps, as shown in **Fig. 3c-1**, which means the repetition rate is 54.2 GHz, corresponding with the sample's FSR. The full width half maximum (FWHM) of each pulse was about 7.8 ps. Such characteristics of the output imply sign of DFWM. Therefore we believe that a certain position of the PCs should exist that can optimize the polarization state of light entering the sample for perfect phase-matching. Thus the PCs were tuned again and more resonant wavelengths began lasing. When the number of lasing lines reached the maximum in our experiment, the oscilloscope's temporal trace showed a 21.6 MHz mode-locking pulse train with the peak intensity declining again, as shown in **Fig. 3a-2, b-2**. Furthermore, **Fig. 3c-2** demonstrates the autocorrelation trace, still a 54.1 GHz pulse sequence with each pulse's FWHM slightly narrowed to 6.6 ps. It's noteworthy that the gigahertz pulse and the Q-switched or mode-locking pulse recorded by oscilloscope are symbionts; once the scope's signal turns to a continuous wave, the autocorrelation trace of high-repetition-rate pulse disappears. Also the mode-locking regime remained unstable and evolved back to the Q-switched mode-locking trace several minutes later with the 54.1 GHz pulse hardly changed. To testify the stability of the fiber laser based on the HPMKR, we kept the laser working steadily for over an hour every time through 6 months; the set-up was still capable of exporting high-repetition-rate pulse train.

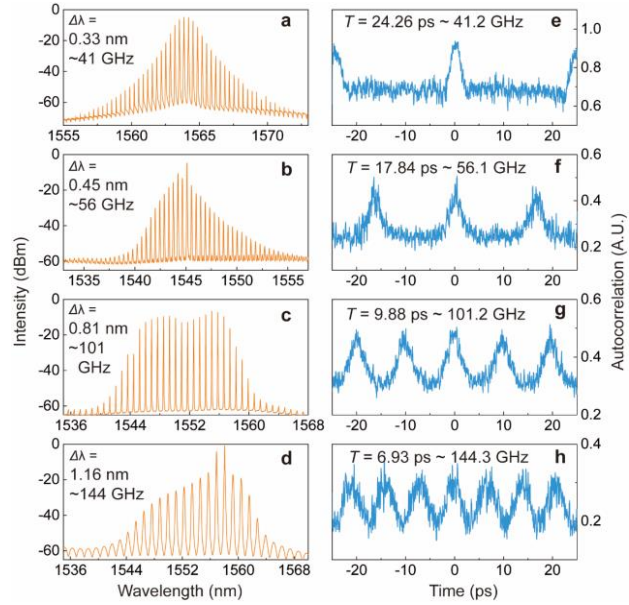

Figure 4 | High-repetition-rate pulse output with different repetition rates.

As mentioned, the sample's FSR is directly linked with the length of MKR's ring section, which can be estimated by $FSR = \lambda^2/(n_{\text{eff}} \times L)$. Considering former experiments that revealed the equivalence between HPMKR's FSR and repetition rate of pulse train, different samples of various sizes were embedded into the laser cavity to verify the DFWM mode-locking mechanism as well as to obtain higher repetition rates. The samples' Q-factors were relatively smaller than that from the former experiment (characterizations of samples' optical properties are provided in **SI**). **Figs. 4a-b** show the optical spectra of laser outputs based on different samples. Each spectrum's corresponding autocorrelation trace was recorded (**Figs. 4e-h**). The HPMKR's FSRs and the pulse train's repetition rates matched well, and a pulse train with a repetition rate as high as 144.3 GHz was achieved.

**Discussion**

As aforementioned three key features of the HPMKR contribute to the output: polarization dependence, filtering, and optical nonlinearity. To better understand the generation mechanism of the gigahertz pulse, comparative trials were performed by detaching the polarizing or nonlinear characteristics from the HPMKR structure. Spectral and temporal measurements of the output of a standard laser based on HPMKR are demonstrated in **Figs. 5a,b** for comparison. Here the HPMKR sample comes with a PDL of 8.1 dB at 1550 nm and insertion loss of 4.55 dB. Data were recorded at the pump of 540 mW, so as to the following comparative experiments. The laser cavity contained a 1.1 m EDF while the total cavity length varied due to different pigtail lengths of the MKR sample or other replacement devices, however, the net dispersions remained anomalous at -0.170, -0.076, -0.081 and -0.173 $ps^2$ respectively.

First we split the polarizing feature from the HPMKR, cascading a PDMS-packaged MKR (with PDL of 0.03 dB which is negligible) with a polarization-dependent isolator (PD-ISO) to displace the HPMKR in the laser cavity. The laser was observed to enter the regime in general the same with that of that of HPMKR laser (**Figs. 5c,d**), although the lasing bandwidth was limited comparatively. Subsequently, the isolator was shifted to a polarization-independent one and all the signals were lost (**Figs. 5e,f**). Such results imply the fundamental role of the HPMKR's polarizing feature in producing stable high-repetition-rate pulses, as well as indicate that the HPMKR's polarization is coupled with the generation process since separation of polarizing and MKR filtering features clearly limited the number of lasing lines and pulse quality. This can be explained by that the polarization state of light passing through PD-ISO still needs adjusting by the PC to satisfy the phase-matching condition of FWM process in the MKR, while it's naturally satisfied in the HPMKR which is both a polarizer and nonlinear element.

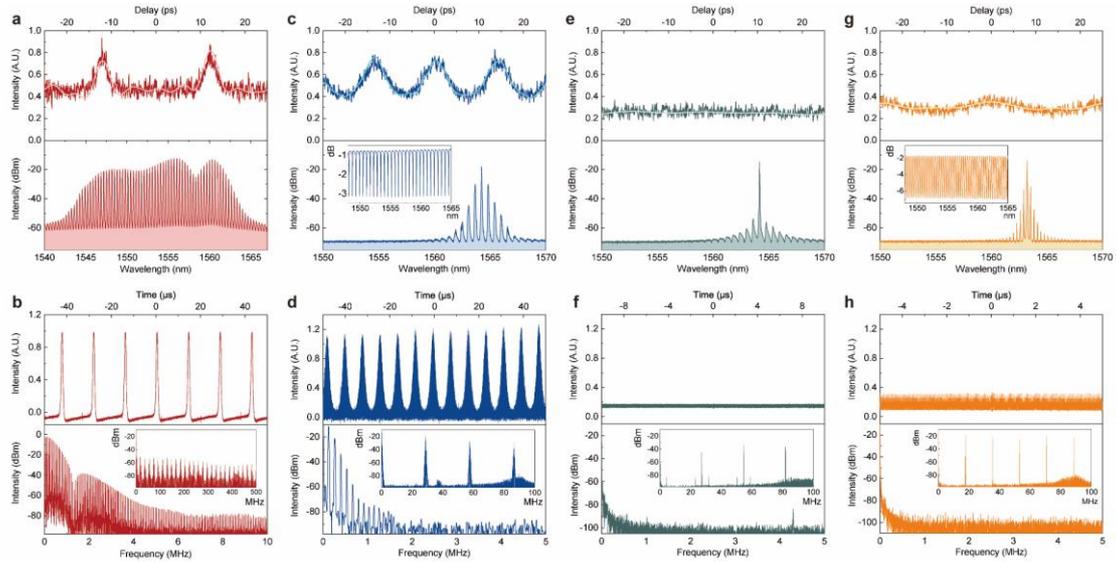

Figure 5 | Laser output of comparative experiments.

To further investigate the function of the microfiber's nonlinearity in laser, we separated this feature by substituting fiber Mach-Zehnder interferometer (MZI) cascaded with a PD-ISO for the HPMKR, where the combined device still exhibited filtering and polarizing properties but the nonlinear gain component vanished. In **Figs. 5g,h**, the measurements indicate no signs of Q-switched mode-locking signals in the typical HPMKR laser or strong NPR mode-locking pulse trains; only weak mode-locked oscilloscope traces hardly distinguished from noise was found, along with narrow-band optical spectra filtered by the FSR of the MZI. The autocorrelation trace showed an evidence of temporal periodic modulation, which can be attributed to faint phase-locking of lasing wavelengths by NPR. Nevertheless regardless of the PC rotation, the trace failed to evolve into a clear pulse sequence. Considering that both PD-ISO and MZI possess significantly better polarizing and filtering performances than most HPMKR samples, it is concluded that the MKR's nonlinearity is important.

According to calculations, the group-velocity dispersion (GVD) and nonlinear coefficient at 1550 nm of the tapered microfiber of diameter about 2.5 μm are -216 ps/(nm·km) and 34.3 $W^{-1}$/km respectively (calculation methods are in **SI**). A high Q-factor and nonlinearity (compared to common high nonlinear fibers) enable DFWM. Additionally, for ring resonators the conditions for frequency-matching and phase-matching are automatically qualified in the HPMKR. The entire laser's operation is analogous to that of lasers based on filter-driven DFWM, where a silica or silicon-nitride micro-ring-resonator is embedded into a loop fiber laser cavity [15, 24]. Since an ultrahigh Q-factor exceeding $10^6$ and large nonlinearity can yield an optical frequency comb (OFC), the scheme is also known as self-locked OFC [25]. As the mode spacing of the laser cavity (typically tens of megahertz) is much smaller than the lasing wavelengths' linewidth (for the HPMKR sample with a Q-factor of $7.9\times10^5$, the linewidth can reach 247 MHz), a number of cavity modes would exist and oscillate in one MKR resonant peak. The beating of cavity modes effectively increases the intracavity peak power to reach the threshold of DFWM but introduces super mode instability as well. Because of the limited recovery time of gain media the temporal trace is modulated periodically by the order of microseconds, and the laser cavity often functions in an unstable Q-switched regime [26, 27, 28].

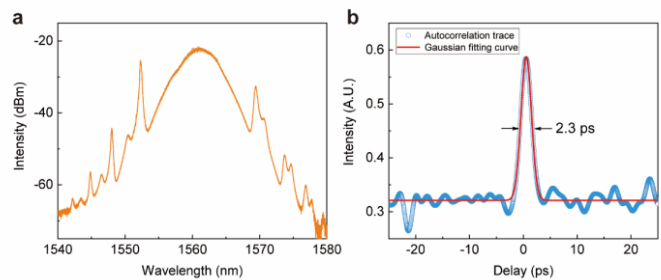

**Figure 6 | NPR mode-locking soliton output of HPMKR laser at low intracavity power.**

But what's in different with FD-DFWM here in laser based on HPMKR is the existence of polarizing feature that brings nonlinear polarization effect to the laser cavity. We placed sample-A into a longer laser cavity, at the pump power of 145 mW, the laser produced soliton NPR-mode-locking pulse train (**Fig. 6**). Other experiments with pump powers less than 150 mW or large-insertion-loss samples yield similar results. Therefore, we conclude that the HPMKR laser tends to function in traditional NPR mode-locking regime at a lower intracavity power. In contrast with arbitrary beating of cavity modes in FD-DFWM, the NPR mode-locking (or NPR Q-switching, depending on the sample's loss and PDL, and cavity length) provides a regular pulse train with stable peak power. Previous comparative experiments confirmed that DFWM cannot be realized without sufficient intracavity power introduced by NPR effect at the beginning. Once the power is increased, the output will evolve to DFWM regime, where super mode instability emerges. Varied with irregular Q-switched pulsations of FD-DFWM [26, 27], the super-mode-instable pulse train in HPMKR laser exhibits a stable Q-switched mode-locking sequence with a standard Gaussian envelope or even mode-locking state owing to modulation from strong saturable absorptions of NPR scheme. Consequently, the stable pulsation offers a larger statistic peak power to enhance DFWM, thereby forming feedback and achieving a steady high-repetition-rate output. In accordance with the analysis above, the DFWM process based on NPR effect in this hybrid plasmonic device shall be considered to be a new mechanism to drive DFWM. We believe that increasing the HPMKR's Q-factor and PDL along with shortening the length of laser cavity would suppress the super mode instability to export high-repetition-rate pulses with better quality.

## Conclusion

In summary, a novel high-repetition-rate mode-locked laser based on a hybrid plasmonic microfiber knot resonator was proposed and demonstrated, and pulses at repetition rates up to 144.3 GHz were generated. Our all-fiber, low-cost, and high-Q device allowed for a new mechanism that enabled stable DFWM mode-locking without long nonlinear elements or costly SOI microresonators, which is termed NPR-stimulated DFWM. In this study the threshold for DFWM mode-locking was lowered, thereby enabling applications of microfiber resonators in laser fields and nonlinear optics, particularly owing to the HPMKR's succinct structure and all-fiber compatibility.

## Methods

**Fabrication of HPMKR.** Beforehand a glass substrate goes through ultrasonic cleaning procedure to avoid the loss introduced by the dust adhering to the glass slide surface. Then a gold film with thickness ~100 nm was deposited on the glass slide by magnetron sputtering. The MKR was made of a silica microfiber that was directly drawn from a standard single-mode optical fiber (SMF-28, Corning, U.S.) by the flame brushing method [29, 30] to a diameter of ~2.5 μm. The knot structure was formed by tying a microfiber by hand with the help of high-precision translation stages [31], which ensures a no-cutting and compact MKR structure made from a double-ended tapered fiber. The MKR was later attached to the prepared glass slide (with gold film) by another translation stage; therefore an HPMKR was realized. To enhance the robustness of the devices, the HPMKR was embedded in PDMS (Sylgard 184 silicone elastomer, Dow Corning). PDMS was selected as the coating material owing to its low refractive index and high optical transparency.

**Measurement of Q-factor.** The Q-factor of the HPMKR samples can be measured using a scanned laser [32]. The output of a wavelength-tunable scanned laser (with a linewidth ~200 kHz) first underwent a PC and a variable optical attenuator, then was transmitted into the sample via a fiber pigtail. A photodetector was employed to receive the transmission signal and the FWHM of the transmitted Lorentz peak was measured using an oscilloscope. The temporal coordinates of the x-axis can be transformed into wavelength domain by the relation

$$\lambda = \lambda_0 + a \cdot b \cdot t$$

Here, $\lambda_0$ is the central wavelength of the scanned laser, $t$ the temporal coordinate, $a$ the piezoelectric coefficient, and $b$ the slope of the triangular wave that modulates the scanning of the laser source. For the measuring system, we set $a$ = 0.041 nm/V and $b$ = 40 V/s.

**Calculation of microfiber's dispersion and nonlinearity.** The group-velocity dispersion of the microfiber is derived from the mode equation of the fiber waveguide[22]

$$\left[\frac{J'_n(u)}{uJ_n(u)} + \frac{K'_n(w)}{wK_n(w)}\right]\left[\frac{J'_n(u)}{uJ_n(u)} + \left(\frac{n_0}{n_1}\right)^2 \frac{K'_n(w)}{wK_n(w)}\right] = n^2\left(\frac{1}{u^2} + \frac{1}{w^2}\right)\left[\frac{1}{u^2} + \left(\frac{n_0}{n_1}\right)^2 \frac{1}{w^2}\right]$$

where $u = v\sqrt{1-b}$, $w = v\sqrt{b}$, $v$ is the normalized frequency and $b$ the normalized propagation constant, $n_1$ and $n_0$ are the refractive indexes of the core and cladding respectively. For the microfiber, the refractive index of cladding PDMS is 1.3997, and the core refractive index of silica is given by Sellmeier polynomial:

$$n(\lambda) = \sqrt{1 + \sum_{i=1}^{3} \frac{a_i \lambda^2}{(\lambda^2 - b_i)}} \quad (\lambda: \mu m)$$

$$\begin{cases} a_1 = 0.6965325 \\ a_2 = 0.4083099 \\ a_3 = 0.8969766 \end{cases} \quad \begin{cases} b_1 = 4.368309 \times 10^{-3} \\ b_2 = 1.394999 \times 10^{-2} \\ b_3 = 9.793399 \times 10^{1} \end{cases}$$

Noting that $n_{\text{eff}} = \sqrt{n_0^2 - (n_1^2 - n_0^2)b}$ and $D_\lambda = -\frac{\lambda}{c} \frac{d^2 n_{\text{eff}}}{d\lambda^2}$, by numerically solving the mode-equation and deriving $b$ we can obtain the dispersion in units of ps/(nm·km). Fig.7a shows the dispersion curves of microfibers of various diameters.

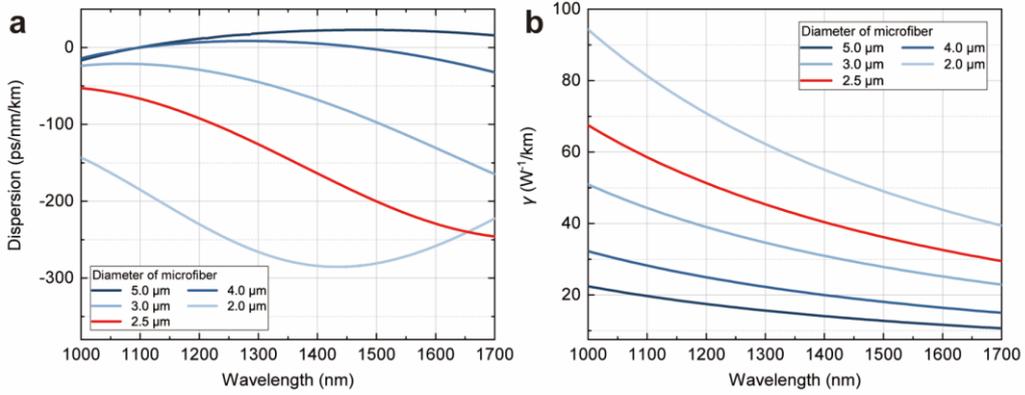

**Figure 7 | Simulation of microfiber's optical parameters.**

The derivation of the microfiber's nonlinear parameter is based on equation [19]

$$\gamma = \frac{2\pi}{\lambda} \frac{\iint_{-\infty}^{+\infty} n_2(x,y)|F(x,y)|^4 dxdy}{\left(\iint_{-\infty}^{+\infty} |F(x,y)|^2 dxdy\right)^2}$$

Here we adopt the nonlinear index $n_2 = 2.5 \times 10^{-20}$ m$^2$/W for silica and 0 for PDMS. The integrations of optical fields were calculated by assuming that the field distributed by a Gaussian profile: $F(x,y) = \exp(-\rho^2/w^2)$. The curves are plotted in Fig. 7b.

**Acknowledgements**

This work was sponsored by the National Key R&D Program of China (2017YFA0303700 and 2017YFA0700503), the National Natural Science Foundation of China (61535005 and 61975107). We thank Prof. J. H. Shun for help in the laser mechanisms, thank Prof. X. S. Jiang and Mr. H. Wang for help in the device measurement.

**Author Contributions**

F.X. and Z.X.D. conceived the experiments. Z.X.D. fabricated and measured the samples and built the laser prototype used in the paper as a proof of concept. Z.X.D., Z.N.H., Y.C., C.B.M., Y.Q.L. and F.X. analyzed the data. Z.N.H. Y.C. and C.B.M. contributed to the laser building and measurement. Y.Q.L. advised on the pulse generation mechanisms. Z.X.D. wrote the original draft. F.X., C.B.M. and Y.Q.L. reviewed and edited the paper. F.X. guided the research and supervised the project.

**Competing Interests**

The authors declare no competing interests

**Figure legends**

1. **HPMKR design and characterization.** (a) Schematic of the central component—a high-Q (Q ~ $10^4$-$10^5$) microfiber ring resonator (with fiber pigtails) attached to gold film and coated with PDMS. Insets show the optical and SEM microscopic pictures of a typical sample that has a ring diameter of 1.1 mm and fiber diameter of 2.4 μm. The red scale bar represents 500 μm. (b, c) Cross-section electric field modal distribution for a TE (LMM) and TM (HMM) polarized beam calculated using the finite element method. The scale bar represents 1.25 μm. Gold and microfiber are denoted in the plots and the ambience is PDMS. (d) Typical polarization dependent loss (PDL) of HPMKR at C-band. The measurement was performed with a 1525-1565 nm ASE source. (e) Exemplary transmission spectra of HPMKR acquired by optical spectrum analyzer and theoretical calculations. The inset provides measurements of this sample's Q-factor up to $7.9 \times 10^5$. The sample labelled sample-A is placed into the laser cavity in the following experiments.

2. **Schematic of the fiber laser based on HPMKR.** The HPMKR sample in **Fig. 1e** is embedded in a loop cavity containing a wavelength division multiplexer (WDM), erbium-doped-fiber (EDF), and polarization-independent isolator (PI-ISO) to force the unidirectional operation, and two polarization controllers to act on the pulse polarization. Furthermore, 20% of the optical waveform was extracted by an 80:20 optical coupler (OC) out of the cavity and monitored by an autocorrelator, optical spectrum analyzer (OSA) and photodetector connected with an oscilloscope as well as an RF spectrum analyzer (RSA).

3. **Evolution process of NPR-stimulated DFWM operation.** a, b and c-series show the experimental optical spectra, oscilloscope traces, and autocorrelation traces of the laser output, respectively for different positions of PCs. To distinguish different working states, we describe each state by the operation regime of oscilloscope traces. (a, b, c) Q-switched operation. (a-1, b-1, c-1) Q-switched mode-locking operation where DFWM mode-locking emerged. (a-2, b-2, c-2) Mode-locking operation, the laser cavity modes were mode-locked by NPR while the HPMKR modes were mode-locked by DFWM. Here, a direct AC trace represents the signal directly captured by autocorrelator and filtered AC trace represents trace filtered by the autocorrelator client's algorithm.

4. **High-repetition-rate pulse output with different repetition rates.** (a-d) Optical spectra of HPMKR laser with FSR-varied samples. The corresponding autocorrelation traces are demonstrated on the right in (e-f).

5. **Laser output of comparative experiments.** (a, b) Output performances of a typical laser based on HPMKR. From top to bottom: the autocorrelation trace, optical spectra, oscilloscope trace, and RF spectra. The inset of the RF spectra is the spectrum under large scale. (c, d) Output of laser with MKR and PD-ISO, the inset of optical spectra shows the MKR's transmission spectra. (e, f) Output of laser with only MKR and PI-ISO (has no polarizing feature). (g, h) Output of laser with MZI (has minimal nonlinearity) and PD-ISO. Inset of optical spectra shows the transmission spectra of MZI.

6. **NPR mode-locking soliton output of HPMKR laser at low intracavity power.** (a) Optical spectra; (b) Autocorrelation trace. The cavity is the same structure as Fig. 2, containing 0.55 m EDF and 12.21 m SMF, corresponding to a net dispersion of -0.245 ps2 and longitudinal mode-spacing of 16.2 MHz.

7. **Simulation of microfiber's optical parameters.** (a) GVD of PDMS cladding microfiber of different diameters; (b) Nonlinear parameter of PDMS cladding microfiber of different diameters.